\documentclass[useAMS,usenatbib,usegraphicx]{mn2e}
\usepackage{times}
\usepackage{aas_macros}
\newcommand{\aX}{\ensuremath{\alpha_{\mathrm{X}}}}
\newcommand{\aUX}{\ensuremath{\alpha^{\mathrm{X}}_{150}}}
\newcommand{\arIR}{\ensuremath{\alpha^{\mathrm{IR}}_{\mathrm{radio}}}}
\title[An optical/X-ray emission component in the jet in
  3C\,273]{Hubble Space Telescope far-ultraviolet imaging of the jet
  in 3C\,273: a common emission component from optical to
  X-rays\thanks{Based on observations made with the NASA/ESA Hubble
    Space Telescope, obtained at the Space Telescope Science
    Institute, which is operated by the Association of Universities
    for Research in Astronomy, Inc., under NASA contract NAS
    5-26555. These observations are associated with HST program
    GO-9814}}

\author[Sebastian Jester et al.]{Sebastian Jester,$^1$\thanks{E-mail:
    jester@mpia.de; this work was begun at the Particle Astrophysics
    Center, Fermilab, Batavia, IL 60510, USA} Klaus Meisenheimer$^1$,
  Andr\'{e} R.\ Martel,$^2$ Eric S.\ Perlman$^3$ \newauthor and William B.\ Sparks$^4$
 \\$^1$Max-Planck-Institut f\"ur Astronomie, K\"onigstuhl
  17, 69117 Heidelberg, Germany
 \\$^2$Department of Physics \& Astronomy, The Johns Hopkins
 University, 3400 N.\ Charles Street, Baltimore, MD 21218, USA 
\\$^3$Physics and Space Sciences Department, Florida Institute of
Technology, 150 West University Boulevard, Melbourne, FL 32901, USA
\\$^4$Space Telescope Science Institute, 3700 San Martin Drive,
Baltimore, MD 21218, USA}
\begin{document}

\date{Accepted 2007 June 17. Received 2007 June 12; in original form 2007 March 23}

\pagerange{\pageref{firstpage}--\pageref{lastpage}} \pubyear{2006}

\maketitle

\label{firstpage}

\begin{abstract}
We present far-ultraviolet (UV) observations at $\sim 150\,$nm of the
jet of the quasar 3C\,273 obtained with the Advanced Camera for
Survey's Solar Blind Channel (ACS/SBC) on board the \emph{Hubble Space
  Telescope}.  While the jet morphology is very similar to that in the
optical and near-ultraviolet, the spectral energy distributions (SEDs)
of the jet's sub-regions show an upturn in $\nu f_\nu$ at 150\,nm
compared to 300\,nm everywhere in the jet.  Moreover, the 150\,nm flux
is compatible with extrapolating the X-ray power-law down to the
ultra-violet region.  This constitutes strong support for a common
origin of the jet's far-UV and X-ray emission. It implies that even a
substantial fraction of the \emph{visible light} in the X-ray
brightest parts of the jet arises from the same spectral component as
the X-rays, as had been suggested earlier based on \emph{Spitzer Space
  Telescope} observations.  We argue that the identification of this
UV/X-ray component opens up the possibility to establish the
synchrotron origin of the X-ray emission by \emph{optical}
polarimetry.
\end{abstract}

\begin{keywords}
quasars: individual: 3C273 -- galaxies:active -- galaxies:jets --
radiation mechanisms: non-thermal -- acceleration of particles
\end{keywords}

\section{Introduction}
\label{s:intro}

The launch of the \emph{Chandra X-ray Observatory} has led to an
order-of-magnitude increase in the number of X-ray detections from
extragalactic jets \citetext{\emph{e.g.},
  \citealp{WBH01,SMTea02,SGMea04,MSLea05}; H.\ Marshall et
  al. \emph{in prep.}}. There is currently a very active debate
about the origin of X-ray emission from many high-power jets, in which
the X-ray and radio emission cannot be explained by a single spectral
component. The X-rays could be due to beamed inverse-Compton (IC)
scattering of cosmic microwave background (CMB) photons, or due to
synchrotron emission from a second electron population
(\citealp{Tav00}; \citealp*{cel00}; \citealp{SSOea04,H06,HK06});
synchrotron self-Compton (SSC) emission from electrons in an
equipartition magnetic field can usually account for the X-ray
emission from \emph{hot spots} at the ends of jets
\citep[see][]{HK06}, but in the jets themselves, SSC is only viable if
the total energy density is dominated by relativistic electrons at the
level of 99.99\% or more, and we will not consider it further here.

One debated case is the jet in 3C273, where data from \emph{ROSAT} and
the first \emph{Chandra} observations had left the X-ray emission
mechanism unclear \citep{Roe00,Mareta01,Sam01}. In \citet{JHMM06}, we
presented deeper Chandra observations of this jet and found that the
X-ray spectra are softer than the radio spectra in nearly all parts of
the jet, ruling out the simplest one-zone beamed IC-CMB (BIC) models.
A BIC model could still work if the jet flow has a spine + sheath (or
otherwise inhomogeneous) structure, but we found synchrotron emission
from a high-energy electron population to be a more plausible
scenario, as velocity shear is very likely to be present and is
capable of accelerating particles \citep{SO02,RD04}.

Evidence for synchrotron X-ray emission in this jet has also come from
an analysis of the overall SED shape.  \citet{Jes02} demonstrated that
all parts of the jet show a UV excess above a simple synchrotron
spectrum fitted to the radio, infrared and optical data, and suggested
that part of the optical/UV emission might be due to the same emission
component as the jet's X-rays.  We therefore proposed to image the jet
at 150\,nm, the shortest UV wavelength accessible with the HST, and
present the results here.

In the meantime, \citet{UUCea06} had analysed Spitzer images at 3.6
and 5.8\,$\umu$m. Surprisingly, they found that the \emph{Spitzer}
fluxes of the first two bright knots A and B1 lie well above the
commonly assumed power-law interpolation between radio and optical/UV.
This indicates that the optical/UV emission is already dominated by
the X-ray component, as suspected by \citet{Jes02}.  Our new results
demonstrate that the emission at 150\,nm is more closely related to
the X-ray than to the radio/IR emission component.

The plan of this paper is as follows: Section \ref{s:obs} describes
the observations and data reduction. In Section \ref{s:res}, we
present the resulting image and SEDs including the new 150\,nm data
point. Section \ref{s:disc} discusses the implications of the new data
for the emission mechanisms and jet structure.

\section{Observations and data reduction}
\label{s:obs}

\begin{table}
\caption{\label{t:obslog}Observation log for new ACS/SBC data. All
  exposures used filter \texttt{F150LP}.}
\begin{tabular}{@{}llll}
\hline
Target & Dataset name & Start time & Exp.\ time \\
 & & UT & s\\
\hline
J122903+020318 & J8P001010 & 04/08/2004 23:13:47 & 900\\
3C273-JET & J8P001TSQ & 04/08/2004 23:32:12 & 1600\\
 3C273-JET & J8P001020 & 05/08/2004 00:45:39 & 2800\\
 3C273-JET & J8P001030 & 05/08/2004 02:21:37 & 2800\\
\hline
\end{tabular}
\end{table}
The Advanced Camera for Surveys (ACS) Solar Blind Channel (SBC) UV
detector was used to image the jet in 3C\,273 through filter
\texttt{F150LP}.  This filter blocks out geocoronal Lyman-$\alpha$
emission and therefore leads to a substantially higher signal-to-noise
ratio than broader filters that extend to shorter wavelengths.  The
total exposure time spent on the jet was 7200\,s. Since the quasar is
too bright in the UV to be imaged with the SBC, we could not use it as
an astrometric reference point.  Therefore, we used 900\,s of our
allocated time to image a nearby star \citep[Star G in][]{RM91} as
absolute astrometric reference point.  Table~\ref{t:obslog} gives the
observation log.

We used the standard pipeline reductions to create flat-fielded
images.  Due to a pipeline bug, information about hot pixels was not
propagated properly to the data quality (DQ) arrays of the
flat-fielded images.  We set the necessary DQ bits for the hot pixels
listed in the appropriate bad-pixel file and then combined all five
individual images using the \texttt{multidrizzle} task from the
\texttt{STSDAS} package, with parameter settings taken from the
pipeline-generated \texttt{mdriztab} table.  The \texttt{multidrizzle}
task corrects for geometric distortion, enabling surface photometry to
be performed on the output image.  No cosmic-ray rejection was
performed, since the SBC is a Multi-Anode Multi-channel Array (MAMA)
which is not sensitive to charged particles.

Visual inspection of the combined image showed that there is a sloping
background, most likely due to a large-angle scattering wing of the
quasar image (the quasar was placed about 7\arcsec\ outside the SBC's
field of view to avoid violating the SBC's count rate limits).  To fit
this background, we used a combination of an exponential profile
representing the scattering wing and a constant level representing the
overall sky background (an exponential by itself, without a constant
offset, left systematic residuals).  As the total background level is
very low, and the relative pixel-to-pixel noise therefore high, the
image was smoothed to 0\farcs2 resolution prior to background
subtraction.  The smoothed image will also be used for all science
analysis.

The absolute pointing of HST data sets can be established only at the
level of 1\arcsec\ \citep[\S 6.2.2]{acs_ihb}, which is clearly not
nearly sufficient for the multi-band multi-telescope analysis of a jet
of width 0\farcs7; however, the absolute roll angle is sufficiently
accurate and precise.  In principle, it is of course possible to use
the brightness peaks of the jet itself to align images at different
wavelengths; however, a spectral index map generated with the help of
such an alignment method is of limited use, because \emph{features in
  spectral index maps arise precisely from morphological differences
  between bands}.  In \citet[Appendix A1]{Jes01} we have shown that
the alignment of images for the creation of a reliable spectral index
map has to be better than 10\% of the PSF width, \emph{i.e.}, at the
level of 20\,mas or better in this case.

As mentioned above, we had planned to use the images of star~G for
absolute astrometric calibration. However, both simply using the
star's position from \citet[i.e., ignoring possible proper
  motions]{RM91}, and correcting of the position using proper motions
from the combined SDSS+USNO-B catalogue by \citet{MMLea04} or the UCAC2
catalogue \citep{ZUZea04} yielded an alignment that left obvious
offsets compared to data taken in other bands. The reason for the
offsets is very likely simply an unhelpful combination of random
errors in the individual positions and in the proper motions. The
coordinates given by \citet{RM91} were obtained roughly 20 years
before the SBC observation; with proper motion random errors of
3\,mas/yr, centroiding errors of 10\,mas (RMS, total for both
coordinates) and an additional contribution from the SBC distortion
correction, the total random error could already be 0\farcs1, or 1/7
of the jet width. This is roughly the magnitude of the observed
offsets.

Hence, absolute astrometry with the accuracy and precision required to
compute a spectral index map is not possible at present.  Therefore,
we restrict the analysis to performing integral photometry of the jet
knots, as in \citet{JHMM06}, for which a visual placement of the
photometry apertures is sufficient.  The photometry was performed in
the same regions shown in Figure~1 of \citet{JHMM06}.

\section{Results}
\label{s:res}

\begin{figure}
\includegraphics[width=84mm]{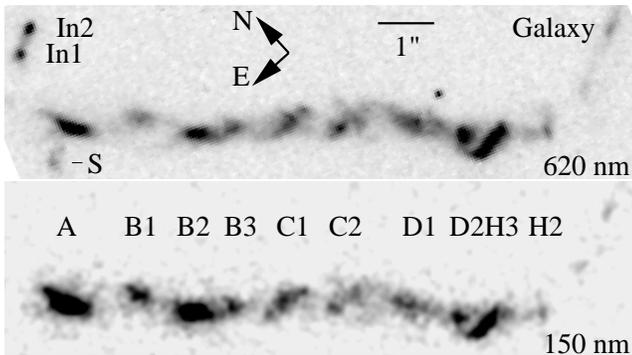}
\caption{\label{f:surfbri}Map of surface brightness (linear grayscale)
  of 3C273's jet at 150\,nm with HST/ACS/SBC/\texttt{F150LP} (bottom),
  and for comparison, at 620\,nm with HST/WFPC2/PC1/\texttt{F622W}
  \protect\citep[top; data from][]{Jes01}.  The images have been
  rotated to place a position angle of $222\degr$ along the
  horizontal. The quasar lies 12\arcsec\ to the northeast of
  knot~A. The names of the regions used in the following figures are
  indicated. On the 620\,nm map, S, In1 and In2 label the ``optical
  extensions'' that are visible at all wavelengths from 1.4\,$\umu$m
  to 300\,nm, but not at 150\,nm (see section
  \protect\ref{s:disc.exts} for a discussion of the extensions).}
\end{figure}
\begin{table*}
\begin{minipage}{124mm}
\caption{\label{t:jetflux}Integrated flux densities of jet regions at
  150\,nm, and spectral indices.  The flux densities are summed over
  the same regions as in \protect\citet{JHMM06}, which extend along
  position angle $222\degr$ over the radial range $r_{\mathrm{in}} \le
  r < r_{\mathrm{out}}$, and to about $\pm0\farcs4$ from the radius
  vector.  Flux densities for all wavelengths are given in
  Tab.~\ref{t:allflux}. The spectral indices are defined as follows:
  $\aX$, spectral index within the \emph{Chandra} band from
  \protect\citet{JHMM06}; $\alpha^{\mathrm{X}}_{\mathrm{150}}$
  spectral index between 1\,keV and 150\,nm; $\arIR$ spectral index
  between 1.4\,$\umu$\,m and 10\,GHz \protect\citep[1.4\,$\umu$m is
    HST NICMOS \texttt{F140W}, 10\,GHz is interpolated from 3 VLA
    bands, all from][]{JRMP05}.}
\begin{tabular}{@{}l*{5}{rr}}
\hline
Region & \multicolumn{1}{c}{$r_{\mathrm{in}}$} &
\multicolumn{1}{c}{$r_{\mathrm{out}}$} &
\multicolumn{1}{c}{$f_{\nu}$} & \multicolumn{1}{c}{$\sigma_f$} &
\multicolumn{1}{c}{$\alpha^{\mathrm{IR}}_{\mathrm{radio}}$} &
\multicolumn{1}{c}{$\sigma(\alpha_{\mathrm{radio}}^{\mathrm{IR}})$} &
\multicolumn{1}{c}{$\alpha^{\mathrm{X}}_{\mathrm{150}}$} &
\multicolumn{1}{c}{$\sigma(\alpha^{\mathrm{X}}_{\mathrm{150}})$} &
\multicolumn{1}{c}{$\alpha_{\mathrm{X}}$} &
\multicolumn{1}{c}{$\sigma(\alpha_{\mathrm{X}})$}\\
& \multicolumn{1}{c}{\arcsec} & \multicolumn{1}{c}{\arcsec} 
& \multicolumn{1}{c}{nJy} & \multicolumn{1}{c}{nJy}  \\
\hline
A	 & 11 & 12.9   & 46.5	& 0.54& -0.90 & -0.05 & -0.78 & -0.01 & -0.83 &  0.02\\
B1	 & 12.9 & 13.8 & 10.9	& 0.25& -0.96 & -0.07 & -0.83 & -0.01 & -0.80 &  0.03\\
B2	 & 13.8 & 14.9 & 20.0	& 0.33& -0.93 & -0.04 & -0.88 & -0.01 & -0.97 &  0.03\\
B3	 & 14.9 & 15.6 & 3.41	& 0.14& -0.93 & -0.09 & -1.00 & -0.02 & -1.13 &  0.07\\
C1	 & 15.6 & 16.5 & 4.85	& 0.16& -0.92 & -0.04 & -1.01 & -0.01 & -1.07 &  0.06\\
C2	 & 16.5 & 17.6 & 6.25	& 0.18& -0.94 & -0.02 & -0.98 & -0.01 & -0.96 &  0.05\\
D1	 & 17.6 & 18.5 & 5.16	& 0.17& -0.95 & -0.02 & -1.00 & -0.01 & -1.02 &  0.05\\
D2H3	 & 18.5 & 20.0 & 7.82	& 0.20& -1.00 & -0.01 & -1.06 & -0.01 & -1.04 &  0.04\\
H2	 & 20.0 & 21.1 & 1.30  & 0.086& -1.22 & -0.00 & -1.11 & -0.02 & -1.27 &  0.12\\
\hline
\end{tabular}
\end{minipage}
\end{table*}

\subsection{Map of surface brightness and photometry of knots}
\label{s:res.map}

Figure~\ref{f:surfbri} shows a map of the surface brightness of the
jet.  The morphology and appearance is identical to that at 620 and
300\,nm \citep[also see Figures 1 and 2 in][]{Jes01}.  The surface
brightness profile continues the trend for the innermost bright knots
to become more dominant at higher frequencies, while the outermost
knots are more dominant at lower frequencies.  However, in a departure
from this trend, the broad edge of H3 appears slightly stronger
compared to the round feature D2 at 150\,nm (peak ratio of about
1.4:1) than at 620\,nm (about 1.25:1).  The resolution of
\emph{Chandra} is not high enough to examine the X-ray morphology at
this level of detail.

\subsection{Spectral energy distributions}
\label{s:res.SED}

\begin{figure*}
\includegraphics[width=0.77\hsize]{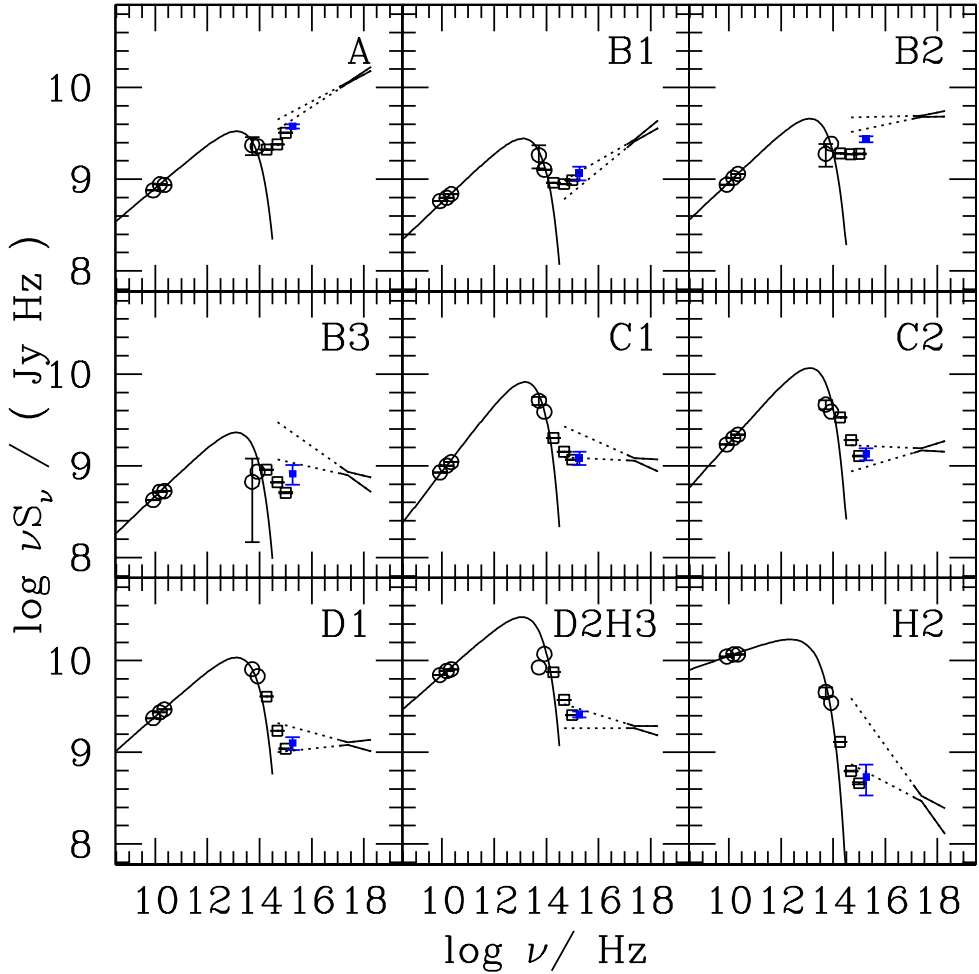}
\caption{\label{f:SEDs}SED of 3C273 jet regions, including VLA,
  Spitzer, HST and Chandra fluxes listed in
  Tab.~\ref{t:allflux}. The new SBC data at 150\,nm are represented
  by solid squares (blue in electronic versions); previous HST data by
  open squares; Spitzer and VLA data by open circles; Chandra data by
  ``bow ties'' (solid lines show the observed region, dotted lines the
  extrapolation into the optical/UV). The solid curves show the an
  exponential cutoff indicating the likely contribution from the
  low-frequency component of the spectrum (these are \emph{not} fits,
  just illustrations, but similar to the curves fitted by
  \citealt{UUCea06}).  In all parts of the jet, the spectral energy
  density at 150\,nm (2$\times 10^{15}$\,Hz) lies above that at
  300\,nm ($10^{15}$\,Hz), but is compatible with lying on or below an
  extrapolation of the Chandra flux to lower frequencies. This is
  strong evidence that the jet emission at 150\,nm is predominantly or
  exclusively due to the same component as the X-rays. In A, B1 and
  B2, even the \emph{optical} emission is dominated by this
  component.}
\end{figure*}
Figure~\ref{f:SEDs} shows SEDs of the jet knots at all presently
available wavelengths, with new SBC measurements reported in
Table~\ref{t:jetflux}.  Observed fluxes were corrected for galactic
extinction using data from \citet*{SFD98} and the galactic extinction
law as given by \citet{Pei92}.  The Spitzer data are taken directly
from \citet[Table 1]{UUCea06}, distributing the reported
\emph{Spitzer} flux density of B2$+$B3 between the individual knots in
the same ratio as observed at 1.4$\umu$m.  

The spatial resolution of \emph{Spitzer} at the wavelengths used, 3.6
and 5.8\,$\umu$m, is 1\farcs66 and 1\farcs88, respectively, which is
much larger than the jet width and larger than, or comparable to, the
knot-to-knot separation.  Therefore, the decomposition by
\citet{UUCea06} of the \emph{Spitzer} flux profile into individual
unresolved components yields measurements that correspond to fluxes
taken with apertures extending about 0\farcs75 to either side of the
jet.  To sample the same spatial regions also at the other
wavelengths, we have used photometry apertures for the VLA and
\emph{HST} data that are considerably larger than those used by
\citet{UUCea06} for the VLA and \emph{HST} data. These larger-aperture
fluxes are up to 30\% higher than those determined by \citet{UUCea06},
with smaller differences for the \emph{HST} measurements of the
fainter knots.  While these differences do not change the overall SED
shape appreciably, we judge the resulting SEDs to be more precise.

The salient feature of Figure~\ref{f:SEDs} is that the spectral flux
per decade ($\nu f_\nu$) at 150\,nm lies \emph{above} that at 300\,nm
in all parts of the jet, but on or below the extrapolation of the
\emph{Chandra} flux and spectral index.

\section{Discussion}
\label{s:disc}

\subsection{Stronger evidence for a common origin of the jet's far-UV and X-ray
  flux}
\label{s:disc.common}

In \citet{Jes02}, we reported an optical/UV excess above a synchrotron
spectrum with a concave cutoff (in $\log f_\nu$ vs.\ $\log \nu$). The
fact that the excess was largest in the X-ray brightest knot suggested
a common origin of the optical/UV excess and the X-rays. Our HST
observations at 150\,nm were designed to clarify whether the X-ray
emission is indeed linked to this excess.  Such a link would have been
ruled out by finding a drop in the SED from 300 to 150\,nm. Instead,
the observed rise of the SED at 150\,nm compared to 300\,nm and the
compatibility of the 150\,nm flux with an extrapolation of the
\emph{Chandra} flux indeed lends strong support to the idea that the
UV excess is due to the same emission component as the X-ray emission.

\begin{figure}
\includegraphics[width=84mm]{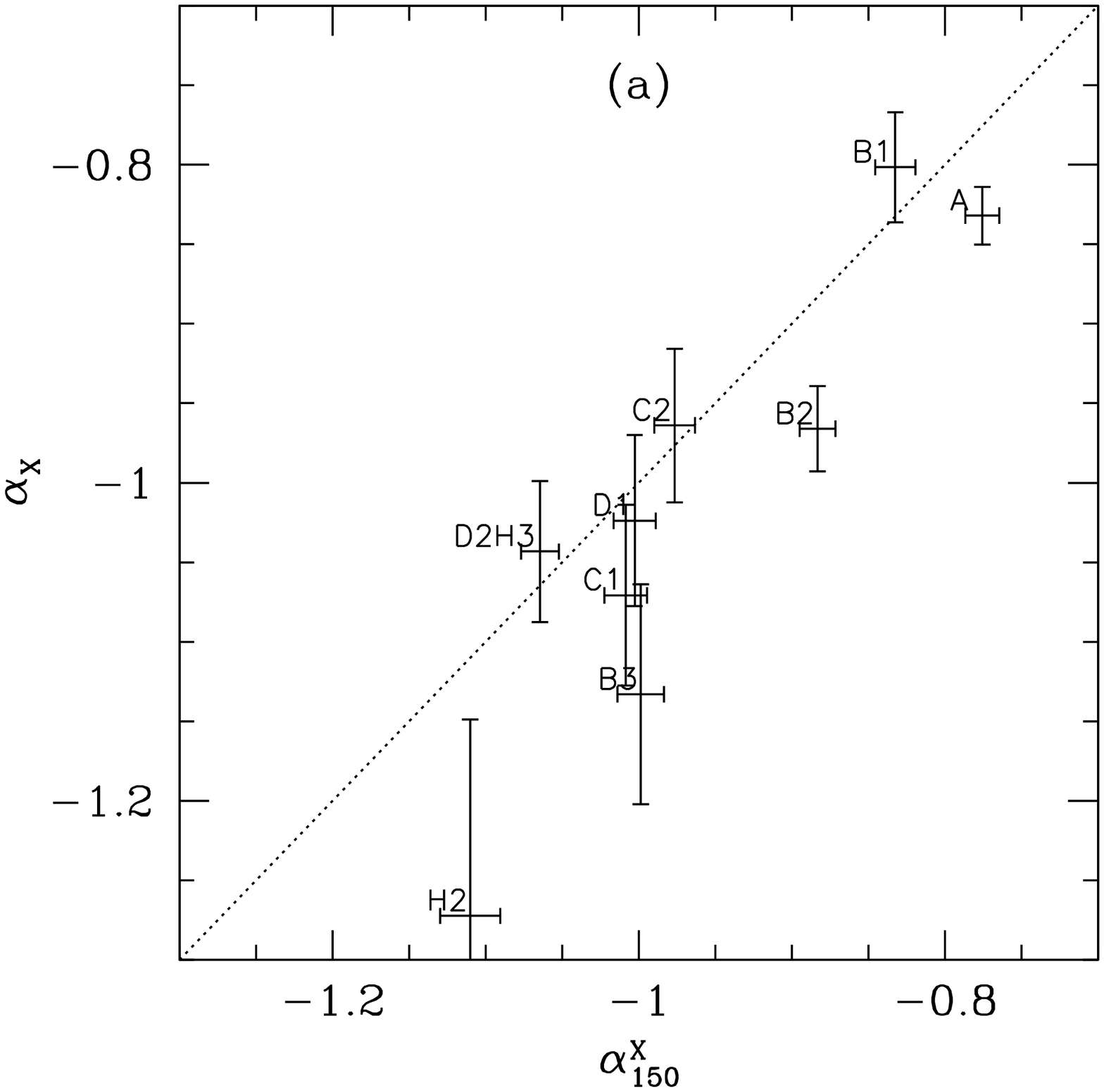}
\includegraphics[angle=270,width=84mm]{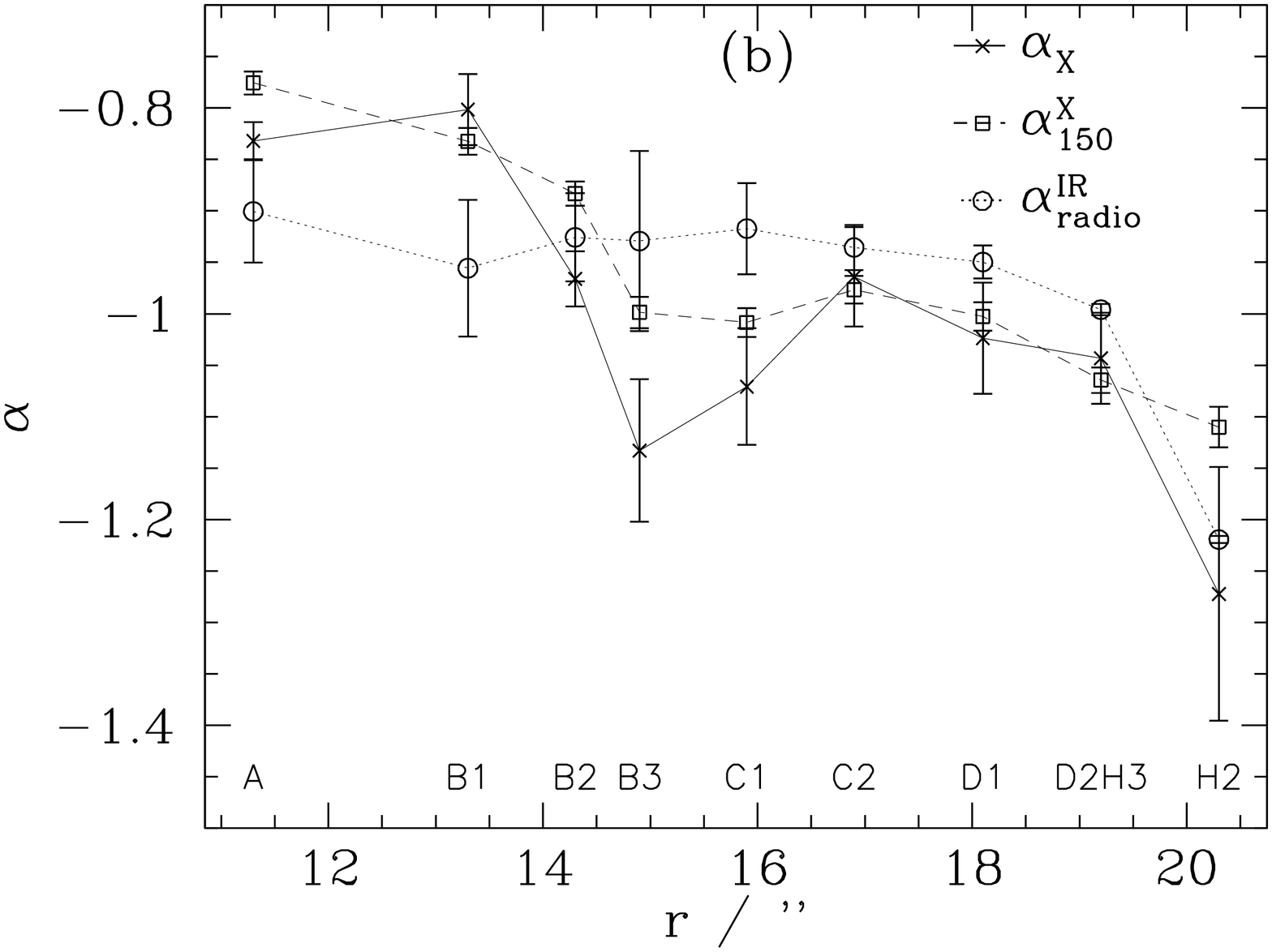}
\caption{\label{f:aUXX}\textbf{(a)} Comparison of $\aX$, the spectral
  index within the \emph{Chandra} band, and $\aUX$, the spectral index
  between 150\,nm and the X-ray flux at 1\,keV (corresponding to
  1.24\,nm and 2.42$\times 10^{17}$\,Hz).  The tight correlation is
  evidence for a common origin of the 150\,nm and X-ray emission. The
  near equality of both spectral indices shows that the UV/X-ray SEDs
  are consistent with power laws (or power laws with a ``cut-on'' in
  the UV region in B2 and B3) whose slopes are evolving along the
  jet. \textbf{(b)} Evolution of spectral indices within the X-ray
  band (0.5-8\,keV, $\aX$), between 150\,nm and X-rays (\aUX) and
  between radio and near-infrared ($1.4\,\umu$m, \arIR) along the
  jet.  Out to $r=18\arcsec$, the observed values of $\arIR$ are
  consistent with being constant at $\arIR=-0.95$, while $\aX$ and
  $\aUX$ drop in accord.}
\end{figure}
\citet{UUCea06} had argued that there is a contribution to the
optical/UV flux from the X-ray component based on the good agreement
between the spectral slope of the high-energy power law they fitted
from optical to X-rays, and $\aX$, the spectral index within the
\emph{Chandra} band.  This is confirmed by the even tighter
correlation between $\aX$ and the far-UV/X-ray spectral index $\aUX$
determined from the SBC and \emph{Chandra} fluxes (Fig.~\ref{f:aUXX}a
and Tab.~\ref{t:jetflux}).  These spectral indices are clearly
consistent with being equal in B1, C1, C2, D1, and D2H3. Due to its
large error bar on $\aX$, the point for in H2 is also compatible with
lying on the equality line, while A, B2 and B3 lie below the equality
line at higher significance. B3 is the smallest region, with an
angular extent similar to the \emph{Chandra} resolution element, and
therefore suffers from the largest systematic uncertainties in
determining spectral indices and it may be an outlier simply due to
these errors.  On the other hand, A and B2 stand out from the
remainder of the jet as the regions with the strongest X-ray emission,
and in the case of A, with an emission peak that is unresolved by
Chandra; this again suggests a difference between these bright regions
and the remainder of the jet, perhaps the presence of a shock that is
responsible for accelerating the X-ray emitting particles.

As pointed out by \citet*{KRA93}, a ``colour-colour'' diagram such as
Figure~\ref{f:aUXX}a is a powerful diagnostic both of the underlying
spectral shape and of the homogeneity of spectral shapes at different
locations in the source.  The similarity of the spectral indices in
Figure~\ref{f:aUXX}a implies that the emission from 150\,nm up to the
\emph{Chandra} band is consistent with being a power law in all parts
of the jet (perhaps with a ``cut-on'', a sharp rise in spectral power
density with increasing frequency, in A, B2 and B3, where $\aX$ dips
below $\aUX$).  Figure~\ref{f:aUXX}b shows that the slope of this
power-law tends to decrease outward along the jet, while the spectral
index $\arIR$ from radio to infrared is consistent with a constant out
to D1 and drops only further out (the constancy of $\arIR$ reflects
the fact that the radio and IR brightness profiles are similar).  In
other words, the spectral shape of the low-energy component remains
constant \citep{Jes01}, while that of the high-energy component
evolves strongly along the jet.

The new SBC data substantially strengthen the conclusion of
\citet{UUCea06} that the jet's far-UV emission is of the same origin
as the X-rays; furthermore, in the X-ray brightest regions A, B1 and
B2, a large fraction of the emission in the optical/near-UV is already
contributed by the same spectral component as the X-rays.  It is
difficult to quantify this fraction exactly, because it depends
sensitively on the detailed shape of the cutoff to the low-energy
emission; we estimate this fraction to be above 50\% in knots A, B1
and B2, but much less in the remainder of the jet.

The fact that the X-ray:radio ratio drops along the jet in the same
way as the X-ray spectral index $\aX$ should provide a hint to the
origin of the X-ray emitting particles. This behaviour is the opposite
of that seen in the so-called \emph{blazar sequence} \citep{FMCea98},
where a lower X-ray flux goes along with a \emph{harder} X-ray
spectrum.  In blazars, this relation is seen as evidence for SSC
emission as origin of the X-rays, which then might argue
\emph{against} an inverse-Compton origin of the X-rays from this jet.

Instead, we speculate that the softening of the X-ray spectrum arises
from re-acceleration of some fraction of the electrons in the
high-energy tail of the ``low-energy'' population.  We ascribe the
spectral softening to a decrease in the acceleration efficiency that
is connected to the increase in the radio:X-ray flux ratio, for
example by changes in the average magnetic field strength.  As
described in \citet{JHMM06}, this re-acceleration could occur in a
thin shear layer around the main jet flow (with the possible exception
of A and perhaps B2, which show evidence for more concentrated X-ray
emission than the remainder of the jet). The variations in radio:X-ray
ratio would be caused both by the decreasing acceleration efficiency,
and by variations in the fraction of high-energy electrons that are
transferred from the radio/optically emitting part of the flow into
the re-acceleration region.  Thus, we consider the ``high-energy''
population to be generated by re-acceleration of electrons initially
in the ``low-energy'' population and escaping into the re-acceleration
region.

Because of the significant contribution of the high-energy spectral
component to the jet flux in the visible wavelength range, we argue
below that it is possible to establish the emission mechanism of the
jet's \emph{X-rays} by performing \emph{optical} polarimetry.  As this
is the first jet to be observed at such short ultraviolet wavelengths,
it is presently not known whether the same is true for other
high-power radio/X-ray jets, though Spitzer observations of
PKS~1136$-$135 suggest a similar SED in its knot~A \citep{UUCea07}.

\subsection{Future observations: optical polarimetry as diagnostic of the X-ray
  emission mechanism}
\label{s:disc.future}

Y.\ Uchiyama \& P.\ Coppi \citetext{\emph{in prep.}} have considered
the polarisation of IC-CMB emission by electrons in a jet with high
bulk Lorentz factor.  We summarize their findings as follows: the
emission from non-relativistic electrons has a linear polarisation,
because the bulk motion of the jet makes the CMB emission anisotropic
in the jet frame, which fixes the scattering geometry and therefore
leads to a net polarisation.  However, electrons with jet-frame
Lorentz factors comparable to, or greater than, the bulk Lorentz
factor can scatter photons from a range of directions into the
observer's line of sight, so that the polarisation is lowered
progressively.  In the case of ultra-relativistic electrons, the bulk
Lorentz factor is negligible compared to the Lorentz factor of the
electrons, so that photons from all azimuthal angles can be scattered
into the observer's line of sight.  Electrons with the mildly
relativistic energies necessary for the production of optical/UV
photons fall into the middle category, with polarisation fraction
around 6\%.  Thus, observation of significantly higher linear
polarisation would rule out the IC-CMB model and instead support a
synchrotron origin for the jet's X-rays.

Unfortunately, the existing polarimetry for this jet has yielded
contradictory results: \citet{RM91} report a polarisation fraction of
$p = 0.07\pm0.04$ based on ground-based data, while \citet*{TMW93}
report $p \approx 0.3$--$0.4$ using the not-yet-aberration-corrected
Faint Object Camera (FOC) on HST.  Also in other parts of the jet, the
ground-based and FOC results are incompatible with each other.  The
ground-based data show the same polarisation fraction and flip in
polarisation angle as radio observations of the radio ``hot spot'',
which is identified with a strong jet-terminating shock
\citetext{\citealp{MH86,magnumopus89}; \citealp*{hs_II}}. Therefore,
we deem the ground-based data to be more reliable and are inclined to
ascribe the discrepancies predominantly to the low signal-to-noise
ratio and the missing spherical aberration correction of the FOC
data. Clearly, additional observations are needed to obtain a
high-fidelity determination of the polarisation fraction, and hence
clarify the X-ray emission mechanism.

\subsection{Why is the jet morphology independent of wavelength even
  with two spectral components?}
\label{s:disc.whymorph}

Even though the brightness profiles of this jet change from radio,
infrared, optical to X-ray wavelengths, the locations of brightness
peaks as well as their angular sizes are strikingly independent of
wavelength (with the exception of B1, whose northern arc is brighter
at infrared and higher frequencies, while the southern one is brighter
at radio frequencies).  This wavelength independence of this jet's
morphology has been somewhat of a puzzle even before it was realised
that there is more than one spectral component. For example, changing
only the magnetic field strength, without varying the shape of the
underlying electron population, should have a much stronger effect on
the optical surface brightness than on that in the radio because the
electron energy distribution is falling more steeply in the optically
emitting range than at low energies.  Nevertheless, the relative
brightness changes from peak to peak are very similar in the radio and
optical.  The puzzle becomes even stronger with the addition of a
second spectral component.

There are further jets showing a similar behaviour, with similar
features at all wavelengths. However, others show strong peaks in one
wavelength region (e.g., X-rays) without a correspondingly strong peak
in another (e.g., radio), \emph{i.e.}, peak shifts indicating a
strongly changing spectral shape \citetext{\emph{e.g.} PKS~1127$-$145
  [\citealp{HK06,SSCea07}] and PKS~1150$+$497
  [\citealp{SGDea06}]}. There are also jets where some knots are like
those in 3C\,273 while others show peak shifts, again indicating a changing
spectral shape, for example the jet in PKS~1136$-$135 \citep{UUCea07}.

For 3C\,273, we speculate (see \S\ref{s:disc.common} above) that the
electron population responsible for the second, high-energy (UV/X-ray)
emission component is fed from the population responsible for the
optical synchrotron emission, and perhaps even caused by the same
process, albeit in different volumes.  By contrast, jets such as
PKS~1127$-$145 may have a different X-ray emission mechanism, a
different mechanism accelerating the X-ray emitting particles, or if
the acceleration is related to velocity shear, differences in their
flow patterns --- though one should bear in mind that the jet lengths
and resolved linear scales differ by large factors between these jets
\citep{HK06}.

\subsection{Nature of ``optical extensions'' to the jet}
\label{s:disc.exts}

Observations at optical wavelengths up to 300\,nm show several
``extensions'' to the jet, whose nature and relation to the jet
material has remained elusive.  Especially the ``southern extension''
$S$ extending towards the southeast from region~A had been speculated
to be related to the jet, both due to its location at the onset of the
jet, and because of its very blue SED between 620 and 300\,nm
\citep[see][Figures~1--3]{Jes01}.  Comparing the optical and
ultraviolet images (Fig.~\ref{f:surfbri}), at 150\,nm the extensions
are conspicuous by their absence.  As the appearance of the jet itself
is not significantly different at 150\,nm from that at longer or
shorter wavelengths, this means that it is unlikely that any of the
extensions consist of the same material and emit by the same mechanism
as the jet itself.  Since the SED of the quasar itself extends well
beyond 150\,nm, too, scattered quasar light is also ruled out as their
origin.  Instead, they are most likely unrelated galaxies.

\section{Summary}
\label{s:sum}

We have used the ACS/SBC on board \emph{HST} with the \texttt{F150LP}
filter to image the jet in 3C\,273 at 150\,nm with the highest
possible resolution.  Our observations show a close agreement of the
jet morphology with that at longer wavelengths (Fig.~\ref{f:surfbri});
the ``optical extensions'' to the jet, however, are undetected at
150\,nm, and we therefore believe that they are unrelated galaxies and
not connected to the jet.  The spectral energy distributions
(Fig.~\ref{f:SEDs}) show that the spectral energy density at 150\,nm
lies above that at 300\,nm in all parts of the jet. The far-UV flux is
compatible with the extrapolation of the 0.5--8\,keV X-ray flux and
spectral index towards lower energies (Figs.~\ref{f:SEDs} and
\ref{f:aUXX}). This lends strong support to a common origin of the
jet's UV and X-ray emission.  This high-energy component is distinct
from the low-energy component responsible for the radio emission.  The
observed rise of the spectral energy density between 300 and 150\,nm
means that the high-energy spectral component does not cut off between
300\,nm and the \emph{Chandra} X-ray band.  This confirms that part of
the jet's emission also at longer wavelengths (620 and 300\,nm) is
already due to the same spectral component as the X-ray emission
\citep[compare][]{UUCea06}.

In the X-ray brightest jet regions, (A, B1 and B2), the optical
emission is even \emph{dominated} by the high-energy component, and
not by the low-energy component accounting for the optical emission in
most of the jet.  Thus, in these regions, optical polarimetry is
equivalent to X-ray polarimetry.  A model explaining the high-energy
component as beamed inverse-Compton scattered cosmic microwave
background (BIC) photons predicts a low degree of linear polarisation
for the optical/UV emission \citetext{Y.\ Uchiyama \& P.\ Coppi,
  \emph{in prep.}}.  Therefore, observing a high degree of
polarisation in these regions would rule out the BIC model for this
jet's X-rays.  The existing polarimetry of 3C\,273 \citep{RM91,TMW93}
has yielded contradictory results, and we have applied to perform
additional polarimetric observations with HST/WFPC2.

In the broader context, it will be important to assess how common it
is for the X-ray component to reach into the optical/UV
region. \emph{Spitzer} observations show similar SEDs as in 3C\,273's
region~A in at least one knot of the jet in PKS 1136$-$135
\citep{UUCea07}.  However, given that 3C\,273 shows a rise in the SED
towards 150\,nm, observations of further jets at this wavelength are
an alternative route to assess in which jets the X-ray spectral
component reaches down to the optical/UV region.  This will enable the
identification of further jets in which optical polarimetry can
determine the X-ray emission mechanism.  In cases where the X-rays are
generated by the BIC mechanism, the low-energy end of the electron
energy distribution can be studied directly in this wavelength region,
while the corresponding radio synchrotron emission emerges at
inaccessibly low frequencies.  Thus, whether synchrotron or
inverse-Compton emission, the low-energy end of the underlying
particle energy distribution can be studied directly around 150\,nm.

\section*{Acknowledgements}

We are grateful to John Biretta for essential advice in planning these
observations, to the staff of the STScI Helpdesk for their dedicated
responses to our requests during the analysis, and to the anonymous
referee for valuable comments.  This work was initiated at the
Fermilab Particle Astrophysics Center with support from the
U.S.\ Department of Energy under contract No.\ DE-AC02-76CH03000.
Support for HST GO program \#9814 at Fermilab was provided by NASA
through a grant from the Space Telescope Science Institute, which is
operated by the Association of Universities for Research in Astronomy,
Inc., under NASA contract NAS 5-26555.  This research has made
extensive use of NASA's Astrophysics Data System Bibliographic
Services.


\appendix

\section{Fluxes for all jet regions}

This appendix gives fluxes for all jet regions collated from previous
works \citep{Jes01,JRMP05,JHMM06,UUCea06} and as plotted in
Figure~\ref{f:SEDs} in a single Table~\ref{t:allflux} for future
reference.  The apertures used for the regions are those shown in
Figure~1 of \citet{JHMM06}.

\begin{table*}
\begin{minipage}{131mm}
\caption{\label{t:allflux}Flux densities for all regions of the 3C273
  jet. Original references by frequency: $8.33\times 10^9, 1.5\times
  10^{10}, 2.25\times 10^{10}$~Hz (VLA) and 1.87$\times 10^{14}$~Hz
  (HST NIC2) from \protect\citet{JRMP05}; 5.23 and 8.45~$\times
  10^{13}$~Hz (\emph{Spitzer} IRAC) from \protect\citet{UUCea06};
  4.85$\times 10^{14}$ and $10^{15}$~Hz (HST WFPC2 \texttt{F622W} and
  \texttt{F300W}) from \protect\citet{Jes01}; 1.86$\times 10^{15}$~Hz
  (HST ACS/SBC) from this work; 2.42$\times 10^{17}$~Hz
  (\emph{Chandra} ACIS-S) from \protect\citet{JHMM06}.}
\begin{tabular}{@{}r*{5}{rr}}
\multicolumn{1}{c}{Frequency [ Hz ]} & $f_\nu$ [ Jy ] & $\sigma_f$ [
  Jy ] & $f_\nu$ [ Jy ] & $\sigma_f$ [ Jy ] & $f_\nu$ [ Jy ] & $\sigma_f$ [ Jy ]\\
\hline
& \multicolumn{6}{l}{Region}\\
& \multicolumn{2}{c}{A} &   \multicolumn{2}{c}{B1} &   \multicolumn{2}{c}{B2} \\
\hline
8.33$\times 10^{9}$  &  0.0905  &  0.0001  &  0.0692  &  8.2$\times 10^{-5}$  &  0.105  &  0.00011 \\
1.5$\times 10^{10}$  &  0.0589  &  5.1$\times 10^{-5}$  &  0.0419  &  4.1$\times 10^{-5}$  &  0.069  &  5.9$\times 10^{-5}$ \\
2.25$\times 10^{10}$  &  0.0385  &  3.8$\times 10^{-5}$  &  0.0306  &  3.2$\times 10^{-5}$  &  0.0508  &  4.6$\times 10^{-5}$ \\
5.23$\times 10^{13}$  &  4.5$\times 10^{-5}$  &  1$\times 10^{-5}$  &  3.5$\times 10^{-5}$  &  1.2$\times 10^{-5}$  &  3.62$\times 10^{-5}$  &  8.1$\times 10^{-6}$ \\
8.45$\times 10^{13}$  &  2.7$\times 10^{-5}$  &  \ldots  &  1.5$\times 10^{-5}$  &  2.6$\times 10^{-6}$  &  2.88$\times 10^{-5}$  &  \ldots \\
1.87$\times 10^{14}$  &  1.13$\times 10^{-5}$  &  1.2$\times 10^{-8}$  &  4.9$\times 10^{-6}$  &  8.3$\times 10^{-9}$  &  1.02$\times 10^{-5}$  &  9.9$\times 10^{-9}$ \\
4.85$\times 10^{14}$  &  4.93$\times 10^{-6}$  &  1.2$\times 10^{-8}$  &  1.83$\times 10^{-6}$  &  8.2$\times 10^{-9}$  &  3.86$\times 10^{-6}$  &  9.8$\times 10^{-9}$ \\
1$\times 10^{15}$  &  3.2$\times 10^{-6}$  &  1.7$\times 10^{-8}$  &  9.73$\times 10^{-7}$  &  1.2$\times 10^{-8}$  &  1.88$\times 10^{-6}$  &  1.3$\times 10^{-8}$ \\
1.86$\times 10^{15}$  &  2.03$\times 10^{-6}$  &  1.1$\times 10^{-7}$  &  6.26$\times 10^{-7}$  &  3.7$\times 10^{-8}$  &  1.47$\times 10^{-6}$  &  7.9$\times 10^{-8}$ \\
2.42$\times 10^{17}$  &  4.65$\times 10^{-8}$  &  5.4$\times 10^{-10}$  &  1.09$\times 10^{-8}$  &  2.5$\times 10^{-10}$  &  2$\times 10^{-8}$  &  3.3$\times 10^{-10}$ \\
\hline
& \multicolumn{2}{c}{B3} &   \multicolumn{2}{c}{C1} &\multicolumn{2}{c}{C2} \\
\hline
8.33$\times 10^{9}$  &  0.0508  &  7.8$\times 10^{-5}$  &  0.101  &  0.00012  &  0.205  &  0.00019 \\
1.5$\times 10^{10}$  &  0.0348  &  4.5$\times 10^{-5}$  &  0.067  &  6.9$\times 10^{-5}$  &  0.134  &  0.00012 \\
2.25$\times 10^{10}$  &  0.0236  &  3.3$\times 10^{-5}$  &  0.0488  &  5.3$\times 10^{-5}$  &  0.0971  &  8.6$\times 10^{-5}$ \\
5.23$\times 10^{13}$  &  1.28$\times 10^{-5}$  &  2.9$\times 10^{-6}$  &  9.8$\times 10^{-5}$  &  1.1$\times 10^{-5}$  &  8.9$\times 10^{-5}$  &  1.2$\times 10^{-5}$ \\
8.45$\times 10^{13}$  &  1.02$\times 10^{-5}$  &  \ldots  &  4.6$\times 10^{-5}$  &  \ldots  &  4.6$\times 10^{-5}$  &  \ldots \\
1.87$\times 10^{14}$  &  4.84$\times 10^{-6}$  &  6.2$\times 10^{-9}$  &  1.08$\times 10^{-5}$  &  8.4$\times 10^{-9}$  &  1.81$\times 10^{-5}$  &  9.9$\times 10^{-9}$ \\
4.85$\times 10^{14}$  &  1.37$\times 10^{-6}$  &  5.9$\times 10^{-9}$  &  2.93$\times 10^{-6}$  &  7.9$\times 10^{-9}$  &  3.93$\times 10^{-6}$  &  9$\times 10^{-9}$ \\
1$\times 10^{15}$  &  5.07$\times 10^{-7}$  &  8.1$\times 10^{-9}$  &  1.16$\times 10^{-6}$  &  1.1$\times 10^{-8}$  &  1.28$\times 10^{-6}$  &  1.2$\times 10^{-8}$ \\
1.86$\times 10^{15}$  &  4.39$\times 10^{-7}$  &  2.8$\times 10^{-8}$  &  6.57$\times 10^{-7}$  &  3.9$\times 10^{-8}$  &  7.23$\times 10^{-7}$  &  4.2$\times 10^{-8}$ \\
2.42$\times 10^{17}$  &  3.41$\times 10^{-9}$  &  1.4$\times 10^{-10}$  &  4.85$\times 10^{-9}$  &  1.6$\times 10^{-10}$  &  6.25$\times 10^{-9}$  &  1.8$\times 10^{-10}$ \\
\hline
&   \multicolumn{2}{c}{D1} & \multicolumn{2}{c}{D2H3} &  \multicolumn{2}{c}{H2} \\
\hline
8.33$\times 10^{9}$  &  0.283  &  0.0003  &  0.836  &  0.00068  &  1.33  &  0.0013\\
1.5$\times 10^{10}$  &  0.182  &  0.00018  &  0.516  &  0.00041  &  0.782  &  0.00074\\
2.25$\times 10^{10}$  &  0.131  &  0.00013  &  0.357  &  0.00028  &  0.52  &  0.00048\\
5.23$\times 10^{13}$  &  1.54$\times 10^{-4}$  &  \ldots  &  1.61$\times 10^{-4}$  &  \ldots  &  8.7$\times 10^{-5}$  &  1.2$\times 10^{-5}$\\
8.45$\times 10^{13}$  &  8$\times 10^{-5}$  &  \ldots  &  1.4$\times
10^{-4}$  &  \ldots  &  4.1$\times 10^{-5}$  &  \ldots \\
1.87$\times 10^{14}$  &  2.17$\times 10^{-5}$  &  9.5$\times 10^{-9}$  &  4$\times 10^{-5}$  &  1.3$\times 10^{-8}$  &  6.97$\times 10^{-6}$  &  9.4$\times 10^{-9}$\\
4.85$\times 10^{14}$  &  3.56$\times 10^{-6}$  &  8$\times 10^{-9}$  &  7.68$\times 10^{-6}$  &  1.1$\times 10^{-8}$  &  1.29$\times 10^{-6}$  &  9$\times 10^{-9}$\\
1$\times 10^{15}$  &  1.09$\times 10^{-6}$  &  1$\times 10^{-8}$  &  2.54$\times 10^{-6}$  &  1.4$\times 10^{-8}$  &  4.59$\times 10^{-7}$  &  1.3$\times 10^{-8}$\\
1.86$\times 10^{15}$  &  6.79$\times 10^{-7}$  &  4$\times 10^{-8}$  &  1.39$\times 10^{-6}$  &  7.5$\times 10^{-8}$  &  2.88$\times 10^{-7}$  &  2$\times 10^{-8}$\\
2.42$\times 10^{17}$  &  5.16$\times 10^{-9}$  &  1.7$\times 10^{-10}$  &  7.82$\times 10^{-9}$  &  2$\times 10^{-10}$  &  1.3$\times 10^{-9}$  &  8.6$\times 10^{-11}$\\
\hline
\end{tabular}
\end{minipage}
\end{table*}

\label{lastpage}

\end{document}